\documentclass[11pt,a4paper]{article}
\pdfoutput=1
\usepackage{jcappub-mod}
\usepackage{ifthen}
\usepackage{epsfig}
\usepackage{booktabs} 
\usepackage{natbib}
\usepackage{bbold}
\DeclareMathAlphabet{\mathbbold}{U}{bbold}{m}{n}   
\usepackage{txfonts}
\usepackage[normalem]{ulem}

\newcommand{\beqra}{\begin{flalign}}
\newcommand{\eeqra}{\end{flalign}}
\newcommand{\beq}{\begin{equation}}
\newcommand{\eeq}{\end{equation}}
\usepackage{ccicons} 
\usepackage{listings}
\usepackage{color}   
\usepackage{braket}  
\usepackage{hyperref}
\usepackage{dsfont}  
\usepackage{leftidx} 
\usepackage{bm}
\DeclareMathAlphabet{\mathpzc}{OT1}{pzc}{m}{it}
\DeclareMathAlphabet{\mathcalligra}{T1}{calligra}{m}{n}

\title{DAMA confronts null searches\\in the effective theory of\\dark matter-nucleon interactions}
\author[a]{Riccardo Catena}
\author[b]{Alejandro Ibarra}
\author[b]{Sebastian Wild}
\affiliation[a]{Chalmers University of Technology, Department of Physics, SE-412 96 G\"oteborg, Sweden}
\affiliation[b]{Physik-Department T30d, Technische Universit\"at M\"unchen,\\James-Franck-Stra\ss{}e, 85748 Garching, Germany}
\emailAdd{catena@chalmers.se}
\emailAdd{ibarra@tum.de}
\emailAdd{sebastian.wild@ph.tum.de}

\abstract{
We examine the dark matter interpretation of the modulation signal reported by the DAMA experiment from the perspective of effective field theories displaying Galilean invariance. We consider the most general effective coupling leading to the elastic scattering of a dark matter particle with spin 0 or 1/2 off a nucleon, and we analyze the compatibility of the DAMA signal with the null results from other direct detection experiments, as well as with the non-observation of a high energy neutrino flux in the direction of the Sun from dark matter annihilation. To this end, we develop a novel semi-analytical approach for comparing experimental results in the high-dimensional parameter space of the non-relativistic effective theory. Assuming the standard halo model, we find a strong tension between the dark matter interpretation of the DAMA modulation signal and the null result experiments. We also list possible ways-out of this conclusion.}

\preprintnumber{TUM-HEP 1036/16}

\begin{document}
\maketitle

\section{Introduction}
\setlength{\parindent}{0.5cm}

The identification of the properties of the dark matter particle is one of the most pressing open questions in Astroparticle Physics. Unfortunately, due to the vastly different characteristics of the many dark matter candidates proposed in the literature, current studies must focus on one single candidate. Among the long list of viable dark matter candidates, the weakly interacting massive particle (WIMP) stands out for its simplicity and its rich phenomenology, which may allow the observation  of dark matter signals in experiments, other than gravitational (for reviews, see~\cite{Bertone:2010zza,Bergstrom00,Jungman:1995df,Bertone:2004pz}).

A promising avenue to detect WIMPs consists in the search for nuclear recoils in a detector induced by their interaction with the dark matter particles that hypothetically permeate our Galaxy~\cite{Goodman:1984dc}. These signals can be detected by the ionization, the scintillation light or the temperature rise in the detector induced by the recoiling nucleus. On the other hand, similar signals can also be generated by electromagnetic interactions of $\alpha$-particles, electrons, and photons produced by the radioactive isotopes in the surrounding material, as well as by nuclear interactions of neutrons produced by natural radioactivity. Therefore, the identification of the very rare dark matter induced recoil events requires a drastic suppression of the backgrounds. The currently most sensitive experiments employ a combination of two detection techniques, which permits to achieve background rates smaller than $1~ \text{event}/(\text{kg}\cdot \text{year})$, which in turn allows to probe the WIMP-nucleus spin-independent interaction cross section at a level better than $10^{-9}~\text{pb}$ for WIMP masses between $\sim 10-100$ GeV~\cite{Akerib:2015rjg}.

An alternative avenue to identify dark matter induced nuclear recoils consists in the search for the characteristic time dependence of the event rate which is expected from a dark matter signal, and which results from the orientation of the Earth orbital plane with respect to the WIMP wind and the consequent different WIMP velocity relative to the Earth over the year~\cite{Drukier:1986tm,Freese:2012xd}. The DAMA experiment, based on radio pure NaI scintillator, and its successor DAMA/LIBRA, have reported an intriguing annually-modulated signal in the single-hit rate in the (2-6) keV energy interval. The modulation has been consistently observed over 14 annual cycles, with a combined significance of 9.3$\sigma$~\cite{Bernabei:2013xsa}. If interpreted in terms of dark matter-induced elastic scatterings, the modulation signal can be reasonably well reproduced for two choices of the dark matter mass and interaction cross-section. Assuming the standard halo model and a purely spin-independent scattering, these are $\sim 11~\text{GeV}$ and $\sim 2\times 10^{-4}~\text{pb}$ (when sodium recoils dominate) and $\sim 76~\text{GeV}$ and $\sim 1.5\times 10^{-4}~\text{pb}$ (when iodine recoils dominate), which are orders of magnitude larger than the upper limits from null searches.  

The tension between the best fit values of the spin-independent cross-section favored by DAMA and the upper limits from other experiments ({\it e.g.} LUX~\cite{Akerib:2013tjd,Akerib:2015rjg}, XENON100~\cite{Aprile:2012nq}, XENON10~\cite{Angle:2011th}, CDMS-Ge~\cite{Ahmed:2010wy}, CDMSlite ~\cite{Agnese:2013jaa}, SuperCDMS~\cite{Agnese:2014aze}, CRESST-II~\cite{Angloher:2014myn}, EDELWEISS-II~\cite{Armengaud:2012pfa}, KIMS~\cite{Kim:2012rza}, XMASS-I~\cite{Abe:2015eos}, SIMPLE~\cite{Felizardo:2011uw}, COUPP~\cite{Behnke:2012ys}, PICASSO~\cite{Archambault:2012pm}, PICO-2L~\cite{Amole:2015lsj,Amole:2016pye} and PICO-60~\cite{Amole:2015pla})  has triggered a vigorous debate about the origin of the modulation signal. It is conceivable that the assumption of dark matter scattering exclusively induced by the spin-independent interaction is too restrictive, and that a consistent picture could arise by allowing other type of interactions. This possibility has been addressed in the literature, see {\it e.g.} \cite{Fitzpatrick:2012ix,DelNobile:2015lxa,Scopel:2015baa,Arina:2014yna,Dolan:2014ska}, however for specific choices or specific combinations of the dark matter-nucleon effective interactions.

In this paper we will perform a systematic exploration of the 28-dimensional parameter space of the Galilean invariant effective theory of elastic dark matter-nucleon interactions~\cite{Fitzpatrick:2012ix}. This is the most general theory for one-body interactions between a dark matter particle and a nucleon, mediated by heavy particles of spin less or equal to 1, and virtually describes the non-relativistic limit of any conceivable model for dark matter-quark or dark matter-gluon interactions~\cite{Anand:2013yka,DelNobile:2013sia,Hill:2013hoa,Catena:2014uqa,Gresham:2014vja,Catena:2014hla,Catena:2014epa,Gluscevic:2014vga,Panci:2014gga,Catena:2015uua,Gluscevic:2015sqa,Dent:2015zpa,Liang:2013dsa,Blumenthal:2014cwa,Catena:2015iea,Catena:2015vpa,Kavanagh:2015jma} fulfilling these basic requirements. We will develop a semi-analytical method that allows to determine  whether a positive signal from one direct detection experiment is incompatible with the null searches, without making assumptions about the nature of the interactions that mediate the dark matter scattering off nuclei. In particular, the method fully takes into account possible interferences among operators, which may lead to suppressed rates in one or in various experiments. We will then apply this method to investigate whether the dark matter interpretation of the DAMA modulation signal is compatible with the null results from other direct detection experiments, and with the non-observation of a high energy neutrino flux in the direction of the Sun from dark matter annihilation. 

This paper is organized as follows: In Section 2 we review the non-relativistic effective theory approach to dark matter-nucleon interactions, in Section 3 we present our method to confront the null results of a given set of experiments with the modulation signal reported by DAMA and in Section 4 we present the numerical results. Lastly, in Section 5 we present our conclusions.

\section{Effective theory of dark matter-nucleon interactions}
\label{sec:eft}
In this section we review the non-relativistic effective theory of one-body dark matter-nucleon interactions.~In doing so, we focus on dark matter scattering from nuclei at fixed target experiments, and in the Sun. 

In the non-relativistic limit, the dark matter scattering amplitude with a nucleon $N$ is restricted by Galilean invariance, {\it i.e.} the invariance under a constant shift of particle velocities, and  momentum conservation.~The most general quantum mechanical Hamiltonian density that generates scattering amplitudes compatible with these restrictions has the following form~\cite{Fitzpatrick:2012ix}:
\begin{eqnarray}
\hat{\bf{\mathcal{H}}}_{\rm N}({\bf{r}})&=& \sum_{\tau=0,1} \sum_{k} c_k^{\tau} \,\hat{\mathcal{O}}_{k}({\bf{r}}) \, t^{\tau} \,,
\label{eq:Hc0c1}
\end{eqnarray}
where $t^{0}= \mathbbold{1}$ is the identity in isospin space, $t^{1}= \tau_3$ is the third Pauli matrix, and ${\mathbf{r}}$ denotes the dark matter-nucleon relative distance.~The Galilean invariant operators $\hat{\mathcal{O}}_k$ in Eq.~(\ref{eq:Hc0c1}) depend on the momentum transfer operator ${\hat{\bf{q}}}$, the transverse relative velocity operator $\hat{\bf{v}}^{\perp}$, and the nucleon and dark matter particle spin operators ${\hat{\bf{S}}}_N$ and ${\hat{\bf{S}}}_\chi$, respectively.~All relevant operators arising for dark matter with spin 0 or 1/2 are listed in Tab.~\ref{tab:operators}. We notice that a spin 1 dark matter particle potentially can induce two additional operators~\cite{Dent:2015zpa}, which we do not include for simplicity, although they might be straightforwardly included in the analysis.~The ``isoscalar'' and ``isovector'' coupling constants in Eq.~(\ref{eq:Hc0c1}), $c_k^0$ and $c_k^1$ respectively, are related to the coupling constants for protons $c_k^p$ and for neutrons $c_k^n$ as follows: $c^{p}_k=(c^{0}_k+c^{1}_k)/2$, and $c^{n}_k=(c^{0}_k-c^{1}_k)/2$.~They have dimension of mass to the power $-2$. The matrix elements of $\hat{\bf{v}}^{\perp}$ are orthogonal to the matrix elements of ${\hat{\bf{q}}}$, which justifies the use of the term ``transverse''.

Under the assumption of one-body dark matter-nucleon interactions, the most general Hamiltonian density that describes the non-relativistic interaction of dark matter with a target nucleus $T$ is given by~\cite{Fitzpatrick:2012ix}
\begin{eqnarray}
\hat{\mathcal{H}}_{\rm T}({\bf{r}}) = \sum_{\tau=0,1}\left[ \hat{\ell}_{\rm V}^{\tau} \hat{\rho}_{\rm V}^{\tau}({\bf r})+ \hat{\ell}_{\rm A}^{\tau} \hat{\rho}_{\rm A}^{\tau}({\bf r}) + \hat{\boldsymbol{\ell}}_{\rm S}^{\tau} \cdot \hat{\boldsymbol{\jmath}}_{\rm S}^{\tau}({\bf r})
+ \hat{\boldsymbol{\ell}}_{\rm C}^{\tau}  \cdot \hat{\boldsymbol{\jmath}}_{\rm C}^{\tau}({\bf r}) + \hat{\boldsymbol{\ell}}_{\rm SV}^{\tau} \cdot \hat{\boldsymbol{\jmath}}_{\rm SV}^{\tau}({\bf r})\right] \,,
\label{eq:HI}
\end{eqnarray}
where the quantum mechanical operators
\begin{eqnarray}
 \hat{\rho}_{\rm V}^{\tau}({\bf r}) &=&  \sum_{i=1}^A \delta({\bf{r}}-{\bf{r}}_i) t_i^{\tau}\,, \nonumber\\
  \hat{\rho}_{\rm A}^{\tau}({\bf r}) &=&  \sum_{i=1}^A  \frac{1}{2m_N} \Bigg[i \overleftarrow{\nabla}_{\bf{r}} \cdot  \boldsymbol{\sigma}_i \delta({\bf{r}}-{\bf{r}}_i) -i \delta({\bf{r}}-{\bf{r}}_i) \
\boldsymbol{\sigma}_i   \cdot  \overrightarrow{\nabla}_{\bf{r}} \Bigg]  t_i^{\tau} \,,
\label{eq:charges}
\end{eqnarray}
are the nuclear vector and axial charges, respectively.~The vectors ${\bf{r}}_i$ and ${\bf{r}}$ are the $i$th-nucleon and dark matter particle position vectors in the nucleus centre of mass frame.~We denote by $\boldsymbol{\sigma}_i$ the 3 Pauli matrices that represent the $i$th-nucleon spin operator. Similarly, $t_i^{\tau}$ is the one-body operator $t^\tau$ defined above now acting on the $i$th-nucleon. The quantum mechanical operators
\begin{eqnarray}
\hat{\boldsymbol{\jmath}}_{\rm S}^{\tau}({\bf r}) &=&  \sum_{i=1}^A \boldsymbol{\sigma}_i \delta({\bf{r}}-{\bf{r}}_i)t_i^{\tau} \,, \nonumber\\ 
\hat{\boldsymbol{\jmath}}_{\rm C}^{\tau}({\bf r}) &=&  \sum_{i=1}^A \frac{1}{2 m_N} \Bigg[i \overleftarrow{\nabla}_{\bf{r}}\delta({\bf{r}}-{\bf{r}}_i) -i \delta({\bf{r}}-{\bf{r}}_i)\overrightarrow{\nabla}_{\bf{r}} \Bigg] t_i^{\tau}\,, \nonumber\\
\hat{\boldsymbol{\jmath}}_{\rm SV}^{\tau}({\bf r}) &=&  \sum_{i=1}^A \frac{1}{2m_N} \Bigg[ \overleftarrow{\nabla}_{\bf{r}} \times \boldsymbol{\sigma}_i  \delta({\bf{r}}-{\bf{r}}_i) +\delta({\bf{r}}-{\bf{r}}_i)\
  \boldsymbol{\sigma}_i  \times \overrightarrow{\nabla}_{\bf{r}} \Bigg] t_i^{\tau} \,,
\label{eq:currents}
\end{eqnarray}
are the nuclear spin, convection and spin-velocity currents, respectively.~The charges and currents in Eqs.~(\ref{eq:charges}) and (\ref{eq:currents}) are model independent, and only reflect our assumption of one-body dark matter-nucleon interactions. 
\begin{table}[t]
    \centering
    \begin{tabular}{ll}
    \toprule
        $\hat{\mathcal{O}}_1 = \mathbbold{1}_{\chi N}$ & $\hat{\mathcal{O}}_9 = i{\hat{\bf{S}}}_\chi\cdot\left(\hat{{\bf{S}}}_N\times\frac{{\hat{\bf{q}}}}{m_N}\right)$  \\
        $\hat{\mathcal{O}}_3 = i\hat{{\bf{S}}}_N\cdot\left(\frac{{\hat{\bf{q}}}}{m_N}\times{\hat{\bf{v}}}^{\perp}\right)$ \hspace{2 cm} &   $\hat{\mathcal{O}}_{10} = i\hat{{\bf{S}}}_N\cdot\frac{{\hat{\bf{q}}}}{m_N}$   \\
        $\hat{\mathcal{O}}_4 = \hat{{\bf{S}}}_{\chi}\cdot \hat{{\bf{S}}}_{N}$ &   $\hat{\mathcal{O}}_{11} = i{\hat{\bf{S}}}_\chi\cdot\frac{{\hat{\bf{q}}}}{m_N}$   \\                                                                             
        $\hat{\mathcal{O}}_5 = i{\hat{\bf{S}}}_\chi\cdot\left(\frac{{\hat{\bf{q}}}}{m_N}\times{\hat{\bf{v}}}^{\perp}\right)$ &  $\hat{\mathcal{O}}_{12} = \hat{{\bf{S}}}_{\chi}\cdot \left(\hat{{\bf{S}}}_{N} \times{\hat{\bf{v}}}^{\perp} \right)$ \\                                                                                                                 
        $\hat{\mathcal{O}}_6 = \left({\hat{\bf{S}}}_\chi\cdot\frac{{\hat{\bf{q}}}}{m_N}\right) \left(\hat{{\bf{S}}}_N\cdot\frac{\hat{{\bf{q}}}}{m_N}\right)$ &  $\hat{\mathcal{O}}_{13} =i \left(\hat{{\bf{S}}}_{\chi}\cdot {\hat{\bf{v}}}^{\perp}\right)\left(\hat{{\bf{S}}}_{N}\cdot \frac{{\hat{\bf{q}}}}{m_N}\right)$ \\   
        $\hat{\mathcal{O}}_7 = \hat{{\bf{S}}}_{N}\cdot {\hat{\bf{v}}}^{\perp}$ &  $\hat{\mathcal{O}}_{14} = i\left(\hat{{\bf{S}}}_{\chi}\cdot \frac{{\hat{\bf{q}}}}{m_N}\right)\left(\hat{{\bf{S}}}_{N}\cdot {\hat{\bf{v}}}^{\perp}\right)$  \\
        $\hat{\mathcal{O}}_8 = \hat{{\bf{S}}}_{\chi}\cdot {\hat{\bf{v}}}^{\perp}$  & $\hat{\mathcal{O}}_{15} = -\left(\hat{{\bf{S}}}_{\chi}\cdot \frac{{\hat{\bf{q}}}}{m_N}\right)\left[ \left(\hat{{\bf{S}}}_{N}\times {\hat{\bf{v}}}^{\perp} \right) \cdot \frac{{\hat{\bf{q}}}}{m_N}\right] $ \\                                                                               
    \bottomrule
    \end{tabular}
    \caption{Non-relativistic Galilean invariant operators for dark matter with spin 0 or 1/2 that are at most linear in the transverse relative velocity operator ${\bf{v}}^{\perp}$, and in nucleon and dark matter particle spin operators, $\hat{\bf{S}}_N$ and $\hat{\bf{S}}_\chi$, respectively.~Operators that are quadratic in ${\bf{v}}^{\perp}$, $\hat{\bf{S}}_N$ and $\hat{\bf{S}}_\chi$ do not arise as leading operators from theories with mediators of spin less or equal to 1~\cite{Anand:2013yka}.~Introducing the nucleon mass, $m_N$, all operators in the table have the same mass dimension, and $q/m_N$ corresponds to the typical internuclear velocity. We label the operators as in~\cite{Anand:2013yka}.} 
    \label{tab:operators}
\end{table}

The model dependent dark matter coupling to the constituent nucleons is described by the quantum mechanical operators $\hat{\ell}_{\rm V}^{\tau}$, $\hat{\ell}_{\rm A}^{\tau}$, $\boldsymbol{\hat{\ell}}_{\rm S}^{\tau}$, $\hat{\boldsymbol{\ell}}_{\rm C}^{\tau}$, and $\hat{\boldsymbol{\ell}}_{\rm SV}^{\tau}$, which exhibit a structure similar to that of the operators in Tab.~\ref{tab:operators}.~The standard spin-independent and spin-dependent interaction operators (in our notation, the operators $\hat{\mathcal{O}}_1$ and $\hat{\mathcal{O}}_4$) only contribute to the operators 
\allowdisplaybreaks
\begin{eqnarray} 
\label{eq:lV}
\hat{\ell}_{\rm V}^{\tau}&=& \left[ c_1^\tau + i  \left( {{\hat{\bf{q}}} \over m_N}  \times {\hat{\bf{v}}}_{T}^\perp \right) \cdot  {\hat{\bf{S}}}_\chi  ~c_5^\tau
+ {\hat{\bf{v}}}_{T}^\perp \cdot {\hat{\bf{S}}}_\chi  ~c_8^\tau + i {{\hat{\bf{q}}} \over m_N} \cdot {\hat{\bf{S}}}_\chi ~c_{11}^\tau \right] 
\end{eqnarray}
and
\begin{eqnarray} 
\hat{\boldsymbol{\ell}}_{\rm S}^{\tau}&=&{1 \over 2} \left[ i {{\hat{\bf{q}}} \over m_N} \times {\hat{\bf{v}}}_{T}^\perp~ c_3^\tau + {\hat{\bf{S}}}_\chi ~c_4^\tau
+  {{\hat{\bf{q}}} \over m_N}~{{\hat{\bf{q}}} \over m_N} \cdot {\hat{\bf{S}}}_\chi ~c_6^\tau
+   {\hat{\bf{v}}}_{T}^\perp ~c_7^\tau + i {{\hat{\bf{q}}} \over m_N} \times {\hat{\bf{S}}}_\chi ~c_9^\tau + i {{\hat{\bf{q}}} \over m_N}~c_{10}^\tau \right. \nonumber \\
  &+&\left.  {\hat{\bf{v}}}_{T}^\perp \times {\hat{\bf{S}}}_\chi ~c_{12}^\tau
+i  {{\hat{\bf{q}}} \over m_N} {\hat{\bf{v}}}_{T}^\perp \cdot {\hat{\bf{S}}}_\chi ~c_{13}^\tau+i {\hat{\bf{v}}}_{T}^\perp {{\hat{\bf{q}}} \over m_N} \cdot {\hat{\bf{S}}}_\chi ~ c_{14}^\tau+{{\hat{\bf{q}}} \over\
 m_N} \times {\hat{\bf{v}}}_{T}^\perp~ {{\hat{\bf{q}}} \over m_N} \cdot {\hat{\bf{S}}}_\chi ~ c_{15}^\tau  \right] \,.
\label{eq:lS}
\end{eqnarray}
In Eqs.~(\ref{eq:lV}) and~(\ref{eq:lS}), ${\hat{\bf{v}}}_{T}^\perp\equiv {\hat{\bf{v}}}^\perp - {\hat{\bf{v}}}_N^\perp$, and ${\hat{\bf{v}}}_N^\perp$ is an operator acting on single-nucleon space coordinates~\cite{Fitzpatrick:2012ix}.~Explicit coordinate space representations for the operators ${\hat{\bf{q}}}$, ${\hat{\bf{v}}}_N^\perp$, and ${\hat{\bf{v}}}_{T}^\perp$ can be found in~\cite{Catena:2015uha}.
In Eq.~(\ref{eq:HI}), the operator
\begin{equation}
\hat{\boldsymbol{\ell}}_{\rm C}^{\tau}=   \left( i {{\hat{\bf{q}}} \over m_N}  \times {\hat{\bf{S}}}_\chi ~c_5^\tau - {\hat{\bf{S}}}_\chi ~c_8^\tau \right) 
\label{eq:lC}
\end{equation}
multiplies the nuclear convention current which, similarly to the nuclear vector charge and spin current operators, is also generated in the case of electroweak scattering from nuclei.~In contrast, the operator 
\begin{equation}
\hat{\boldsymbol{\ell}}_{\rm SV}^{\tau}= {1 \over 2} \left[  {{\hat{\bf{q}}} \over m_N} ~ c_3^\tau +i {\hat{\bf{S}}}_\chi~c_{12}^\tau - {{\hat{\bf{q}}} \over  m_N} \times{\hat{\bf{S}}}_\chi  ~c_{13}^\tau-i {{\hat{\bf{q}}} \over  m_N} {{\hat{\bf{q}}} \over m_N} \cdot {\hat{\bf{S}}}_\chi  ~c_{15}^\tau \right] 
\label{eq:lSV}
\end{equation}
and the nuclear spin-velocity current are specific to dark matter-nucleon interactions.~Here we neglect the operator $\hat{\ell}_{\rm A}^{\tau}$, since the nuclear axial charge operator does not contribute to scattering cross-sections for nuclear ground states that are eigenstates of P and CP, and we assume that this is the case for the nuclei considered here.   
 
From the Hamiltonian density in Eq.~(\ref{eq:HI}), one can calculate the differential cross-section for non-relativistic dark matter scattering from target nuclei of mass $m_T$ and spin $J$. It is given by
\begin{equation}
\frac{{\rm d} \sigma_T(v^2,E_R)}{{\rm d} E_R} = \frac{m_T}{2\pi v^2} \,\langle |\mathcal{M}_{\rm NR}|^2\rangle_{\rm spins}\,,
\label{eq:sigma}
\end{equation} 
and can non-trivially depend on the dark matter-nucleus relative velocity $v\equiv|\vec{v}|$, and on the nuclear recoil energy $E_R$.~The spin averaged square modulus of the non-relativistic scattering amplitude is given by~\cite{Fitzpatrick:2012ix}
\begin{align}
\langle |\mathcal{M}_{\rm NR}|^2\rangle_{\rm spins} = \frac{4\pi}{2J+1} \sum_{\tau,\tau'} &\bigg[ \sum_{k=M,\Sigma',\Sigma''} R^{\tau\tau'}_k\left(v_T^{\perp 2}, {q^2 \over m_N^2} \right) W_k^{\tau\tau'}(q^2) \nonumber\\
&+{q^{2} \over m_N^2} \sum_{k=\Phi'', \Phi'' M, \tilde{\Phi}', \Delta, \Delta \Sigma'} R^{\tau\tau'}_k\left(v_T^{\perp 2}, {q^2 \over m_N^2}\right) W_k^{\tau\tau'}(q^2) \bigg] \,.
\label{eq:M2}
\end{align}  
Eq.~(\ref{eq:M2}) is the sum of eight terms.~Each term is the product of a nuclear response function $W_k^{\tau\tau'}$ and a dark matter response function $R_k^{\tau\tau'}$ (here we follow the same notation introduced in~\cite{Fitzpatrick:2012ix}). The nuclear response functions $W_k^{\tau\tau'}$ are quadratic in matrix elements reduced in the spin magnetic quantum number of the nuclear charges and currents.~In this work, we use the nuclear response functions derived in~\cite{Anand:2013yka} to evaluate scattering rates at dark matter direct detection experiments, and the nuclear response functions found in~\cite{Catena:2015uha} to compute the rate of dark matter capture by the Sun.

The dark matter response functions $R_k^{\tau\tau'}$ are analytically known, and depend on $q^2$ and $v_T^{\perp 2}=v^2-q^2/(4 \mu_T^2)$, where $\mu_T$ is the dark matter-nucleus reduced mass.~They are quadratic in matrix elements of the operators in Eqs.~(\ref{eq:lS}), (\ref{eq:lV}), (\ref{eq:lC}) and (\ref{eq:lSV}).~We list them in appendix~\ref{sec:appDM}. Inspection of Eq.~(\ref{eq:R}) shows that isoscalar and isovector components of a given operator $\hat{\mathcal{O}}_i$ interfere, because of terms proportional to $c_i^\tau c_i^{\tau'}$, $\tau\neq\tau'$, in the differential cross-section.~We also find that seven pairs of operators interfere, because of terms proportional to $c_i^\tau c_j^{\tau'}$, $i\neq j$, in Eq.~(\ref{eq:sigma}).~Here the indexes $i$ and $j$ identify the following pairs of operators:~$(\hat{\mathcal{O}}_1,\hat{\mathcal{O}}_3)$, $(\hat{\mathcal{O}}_4,\hat{\mathcal{O}}_5)$, $(\hat{\mathcal{O}}_4,\hat{\mathcal{O}}_6)$, $(\hat{\mathcal{O}}_8,\hat{\mathcal{O}}_9)$, $(\hat{\mathcal{O}}_{11},\hat{\mathcal{O}}_{12})$, $(\hat{\mathcal{O}}_{11},\hat{\mathcal{O}}_{15})$, and $(\hat{\mathcal{O}}_{12},\hat{\mathcal{O}}_{15})$.~Other pairs of operators do not interfere, partially since nuclear ground states are assumed to be eigenstates of P and CP, and partially because of their $ {\bf{\hat{S}}}_\chi$ dependence.~For a detailed discussion on the impact of similar interference patterns in the calculation of dark matter direct detection exclusion limits, see~\cite{Catena:2015uua}.     

From the differential scattering cross-section in Eq.~(\ref{eq:sigma}), one can compute the differential event rate per unit time and per unit detector mass expected at a direct detection experiment:
\begin{equation}
\frac{{\rm d}\mathcal{R}_T}{{\rm d}E_{R}} =  
 \xi_T \frac{\rho_{\chi}}{m_\chi m_T}  
 \int_{v > v_{\rm min}(q)} \, f(\vec{v} + \vec{v}_{e}(t)) v \, \frac{{\rm d} \sigma_T(v^2,E_R)}{{\rm d} E_R}\, d^3v \,,
\label{eq:diffrate_theory}
\end{equation}
where $\xi_T$ is the mass fraction of the nucleus $T$ in the target material, $m_\chi$ is the dark matter particle mass, and $\rho_\chi$ is the local dark matter density.~In Eq.~(\ref{eq:diffrate_theory}), $f$ is the local dark matter velocity distribution in the galactic rest frame boosted to the detector frame, $\vec{v}_e(t)$ is the time-dependent Earth velocity in the galactic rest frame, and $v_{\rm min}(q)=q/2\mu_T$ is the minimum velocity required to transfer a momentum $q$ in the scattering.~Here we consider a Maxwell-Boltzmann distribution $f(\vec{v} + \vec{v}_{e}(t))\propto \exp(-|\vec{v}+\vec{v}_{e}(t)|^2/v_0^2)$ truncated at the local escape velocity $v_{\rm esc}$. For $v_0$ and $v_{\rm esc}$ we will consider a sample of reference values, as we will see below. Finally, the total number of recoil events expected at a direct detection experiment is $N=\mathcal{D}\cdot R$, with $\mathcal{D}$ the exposure of the experiment and $R$ the event rate, which follows from
\begin{equation}
  R =  \int_0^\infty \text{d}E_R \, \sum_{T} \epsilon_T(E_R) \, \frac{{\rm d}\mathcal{R}_T}{{\rm d}E_{R}} \;,
 \label{eq:totalrate_theory}
\end{equation}
where $\epsilon_T(E_R)$ is the probability that a nuclear recoil off the target nucleus $T$ with energy $E_R$ is detected.

The DAMA experiment, on the other hand, searches for the time modulation of the nuclear recoil event rate. The modulation amplitude in a given energy bin specified by the upper and lower boundaries $E_-$ and $E_+$, respectively, is defined by:
\begin{align}
S_{\text{DAMA}[E_-,E_+]} = \frac{1}{E_{+}-E_{-}} \cdot \frac12 \cdot \left( R_{\text{DAMA}[E_-,E_+]}\Big|_\text{June 1st} - R_{\text{DAMA}[E_-,E_+]}\Big|_ \text{Dec 1st} \right) \,,
\label{eq:modulation-DAMA}
\end{align}
where $R_{\text{DAMA}[E_-,E_+]}(t)$ is the total event rate in that bin at the time $t$, which can be calculated from Eq.~(\ref{eq:totalrate_theory}), using the efficiency $\epsilon_T^{\text{DAMA}[E_-,E_+]}(E_R) = \Phi \left( Q_T E_R, E_{-}, E_{+}\right)$. Here, $Q_T$ is the quenching factor for the isotope $T\in \left\{ \text{Na},\text{I}\right\}$, and $\Phi (Q_T E_R,E_{-}, E_{+})$ is the probability that an event with a nuclear recoil energy $E_R$ (and hence with a quenched energy of $Q_T E_R$) is detected in the energy bin $[E_{-}, E_{+}]$. For that, we assume a Gaussian energy resolution as specified in~\cite{Savage:2008er}.

Another avenue to probe the dark matter-nucleon interaction consists in the search for a high energy neutrino flux correlated to the direction of the Sun, hypothetically produced by the annihilation of dark matter particles which have been previously captured in the solar core by the above-mentioned interactions.  The dark matter capture rate in the Sun reads:
\begin{equation}
\frac{{\rm d} C}{{\rm d}V} = \int_{0}^{\infty} {\rm d}u\, \frac{\tilde{f}(u)}{u}\, w\Omega_{v}^{-}(w) \,,
\label{eq:drate}
\end{equation}
with $\tilde{f}$ given by a truncated Maxwell-Boltzmann distribution integrated over the angular variables.~The rate of scattering from a velocity $w$ to a velocity less than the escape velocity $v(R)$ at a distance $R$ from the Sun's centre is given by~\cite{Gould:1987ir}
\begin{equation}
\Omega_{v}^{-}(w)= \sum_T n_T w\,\Theta\left( \frac{\mu_T}{\mu^2_{+,T}} - \frac{u^2}{w^2} \right)\int_{E u^2/w^2}^{E \mu_T/\mu_{+,T}^2} {\rm d}E_R\,\frac{{\rm d}\sigma_{T}\left(w^2,E_R\right)}{{\rm d}E_R}\,,
\label{eq:omega}
\end{equation}
where $E=m_\chi w^2/2$ is the dark matter initial kinetic energy, $n_T(R)$ is the density of the target nuclei $T$ at $R$, and $u$ is the dark matter velocity in the rest frame of the Sun at $R\rightarrow \infty$, where the Sun's gravitational potential is negligible.~Since $w=\sqrt{u^2+v(R)^2}$, the rate $\Omega_{v}^{-}(w)$ depends on the radial coordinate $R$. The sum in the scattering rate (\ref{eq:omega}) extends over the most abundant elements in the Sun, and the dimensionless parameters $\mu_T$ and $\mu_{\pm,T}$  are defined as follows 
\begin{equation}
\mu_T\equiv \frac{m_\chi}{m_T}\,; \qquad\qquad \mu_{\pm,T}\equiv \frac{\mu_T\pm1}{2}\,.
\end{equation}

In summary, eqs.~(\ref{eq:totalrate_theory}), (\ref{eq:modulation-DAMA}), (\ref{eq:drate}) can be used to calculate the event rates in direct detection experiments and neutrino telescopes for a given underlying particle physics model of dark matter. In the framework of the non-relativistic effective theory, the latter is fully specified by the dark matter mass and the set of the 28 coefficients $c_k^{\tau}$ corresponding to the interaction terms in eq.~(\ref{eq:Hc0c1}).  More specifically, for any of the experiments discussed in this paper, the event rate can be written in the form
\begin{align}
\textrm{event rate} \, \propto \, {\bf c}^T \varmathbb{X}  \;{\bf c} \,.
\label{eq:quadratic_dependence}
\end{align}
Here, ${\bf c}=\left(c_1^{(0)}, c_1^{(1)},c_3^{(0)}, c_3^{(1)},..., c_{15}^{(0)}, c_{15}^{(1)}\right)^T$ is a 28-dimensional vector specifying the concrete particle physics model, while $\varmathbb{X}$ is a real symmetric $28 \times 28$ matrix, which encodes all the information about nuclear responses, the dark matter velocity distribution, experimental efficiencies, etc., but which is \emph{independent} of the underlying particle physics model (for a given dark matter mass). As we will show in the next section, this factorization allows to efficiently compare the results of various direct detection experiments and neutrino telescopes in the high-dimensional parameter space of the non-relativistic effective theory, without making any a priori assumptions regarding the relative size of the various Wilson coefficients $c_k^\tau$.

\section{Confronting the DAMA signal with null result experiments}
\label{sec:Confrontation} 

The DAMA collaboration has reported evidence for the annual modulation of the scintillation light in sodium iodine detectors, which has been interpreted as the result of the time-dependent interaction rate of dark matter particles with the nuclei in the detector due to the orbital motion of the Earth around the Sun~\cite{Bernabei:2013xsa}. More concretely, the modulation amplitudes, as defined in Eq.~(\ref{eq:modulation-DAMA}), measured in the energy bins [2.0, 2.5], [2.5, 3.0] and [3.0, 3.5] keV are, respectively,  $(1.75\pm 0.37)\times 10^{-2}$, $(2.51 \pm 0.40)\times 10^{-2}$ and $(2.16 \pm 0.40)\times 10^{-2}~\text{day}^{-1}\,\text{kg}^{-1} \, \text{keV}^{-1}$.

The modulation signal can, in principle, be explained by elastic dark matter scatterings induced by any combination of the non-relativistic operators $\hat {\cal O}_i$ introduced in Section \ref{sec:eft}. Following Eq.~(\ref{eq:quadratic_dependence}), the  $n_\sigma$-significance lower limit (l.l.) on the modulation amplitude in the bin $[E_-,E_+]$ can be cast as 
\begin{equation}
{\bf c}^T \; \varmathbb{S}_\text{DAMA}(E_-,E_+;m_\chi)  \;{\bf c}> S^{n_\sigma\text{-l.l.}}_{\text{DAMA}[E_-,E_+]}\;,
\label{eq:ll-DAMA}
\end{equation}
with ${\bf c}=\left(c_1^{(0)}, c_1^{(1)},c_3^{(0)}, c_3^{(1)},..., c_{15}^{(0)}, c_{15}^{(1)}\right)^T$ and $\varmathbb{S}_\text{DAMA}(E_-,E_+,m_\chi)$ a $28\times 28$ matrix which depends only on the chosen energy bin and on the dark matter mass, and which can be calculated from Eq.~(\ref{eq:modulation-DAMA}). Analogously, the $n_\sigma$-significance upper limit (u.l.) on the modulation amplitude is
\begin{equation}
{\bf c}^T \; \varmathbb{S}_\text{DAMA}(E_-,E_+;m_\chi)  \;{\bf c}< S^{n_\sigma\text{-u.l.}}_{\text{DAMA}[E_-,E_+] }\;.
\label{eq:ul-DAMA}
\end{equation}

On the other hand, several direct detection experiments have set an upper limit on the total number of recoil events which, in analogy to Eq.~(\ref{eq:ul-DAMA}), can be cast as
\begin{equation}
{\bf c}^T \; \varmathbb{N}_j(m_\chi) \;{\bf c}< N^{n_\sigma\text{-u.l.}}_j\;,
\label{eq:ul-DD}
\end{equation}
where the matrices $\varmathbb{N}_j(m_\chi)$ can be calculated from Eq.~(\ref{eq:totalrate_theory}) and depend on the dark matter mass and on the experimental set-up, which we label by $j$. In our analysis, we will use the null results from LUX~\cite{Akerib:2015rjg}, SuperCDMS~\cite{Agnese:2014aze}, SIMPLE~\cite{Felizardo:2011uw}, PICO~\cite{Amole:2016pye}, COUPP~\cite{Behnke:2012ys} (with nucleation thresholds $E_T=7.8$, 11.0 and 15.5 keV) and PICASSO~\cite{Archambault:2012pm} (with nucleation thresholds $E_T=1.73$ and 2.9 keV); the corresponding 95\% and 99.9\% C.L. upper limits on the number of recoil events are listed in Tab.~\ref{tab:ul-ll}.\footnote{Details about the derivation of the upper limits can be found in Appendix~\ref{sec:DD_analysis_details}.}

\begin{table}
\begin{center}
\begin{tabular}{ | l || l | l |}
  \hline
   Experiment & $N^{95\%\text{-u.l.}}$ & $N^{99.9\%\text{-u.l.}}$ \\ \hline \hline
  LUX & 4.74 events & 9.23 events\\
  SuperCDMS & 18.2 events & 25.6 events\\
  SIMPLE & 5.26 events & 11.9 events\\
  PICO & 4.74 events & 9.23 events\\
  COUPP ($E_T=7.8$ keV) & 6.30 events & 11.2 events\\
  COUPP ($E_T=11$ keV) & 7.75  events & 13.0 events\\
  COUPP ($E_T=15.5$ keV) & 14.4 events & 21.1 events\\
  PICASSO ($E_T=1.73$ keV) & $8.72\,\text{events}/ (\text{kg}_F\cdot\text{day})$ & $18.0\,\text{events}/ (\text{kg}_F\cdot\text{day})$\\
  PICASSO ($E_T=2.9$ keV) & $3.21\,\text{events}/ (\text{kg}_F\cdot\text{day})$ & $5.59\,\text{events}/ (\text{kg}_F\cdot\text{day})$\\
  \hline  
\end{tabular}
\end{center}
\caption{$95 \%$ and $99.9 \%$ C.L. upper limits on the number of dark matter-induced scattering events for the set of experiments selected in our analysis (for details, see  appendix~\ref{sec:DD_analysis_details}).}
\label{tab:ul-ll}
\end{table}

The null results from direct detection experiments are complemented by the null results from neutrino telescopes, which have not observed a significant excess in the high energy neutrino flux in the direction of the Sun. Under the assumption that dark matter capture and annihilation are in equilibrium, the non-observation of the high energy neutrinos hypothetically produced in the dark matter annihilation imply, for a given dark matter mass and annihilation channel, an upper limit on the capture rate and hence an upper limit on the dark matter scattering rate with the nuclei in the solar interior. Using again the notation of Eq.~(\ref{eq:quadratic_dependence}), the $n_\sigma$-upper limit reads
\begin{equation}
{\bf c}^T \; \varmathbb{C}_j(\chi\chi\rightarrow \text{final};m_\chi)  \;{\bf c}< C^{n_\sigma\text{-u.l.}}_j(\chi\chi\rightarrow \text{final};m_\chi)\;,
\label{eq:ul-SK}
\end{equation}
where the matrix $\varmathbb{C}_j(\chi\chi\rightarrow \text{final};m_\chi) $  can be calculated from Eq.~(\ref{eq:drate}), $j$ labels the neutrino telescope and ``final''  denotes the final state in the annihilation. Concretely, we will use the null results from Super-Kamiokande, and we will consider, following this experiment, the final states $b\bar b$ and $\tau^+\tau^-/W^+W^-$, which are archetypes of a soft and a hard neutrino spectrum, respectively. The 90\% C.L. upper limits on the capture rate as a function of the dark matter mass can be found in~\cite{Choi:2015ara}, which we translate into 95\% and 99.9\% C.L. limits, assuming that the corresponding likelihood has a gaussian distribution centered at zero signal.

For our analysis we find convenient to define the rescaled matrices
\begin{align}
&
\varmathbb{A}^{n_\sigma\text{-l.l}}_{\text{DAMA}[E_-,E_+]}\, (m_\chi)  \equiv \frac{\varmathbb{S}_\text{DAMA}(E_-,E_+;m_\chi)}{S^{n_\sigma\text{-l.l}}_{\text{DAMA}[E_-,E_+]} }\;, & & 
\varmathbb{A}^{n_\sigma\text{-u.l.}}_{\text{DAMA}[E_-,E_+]}\, (m_\chi)  \equiv \frac{\varmathbb{S}_\text{DAMA}(E_-,E_+;m_\chi)}{S^{n_\sigma\text{-u.l.}}_{\text{DAMA}[E_-,E_+]}\,}\;,
\nonumber \\
& \varmathbb{A}^{n_\sigma\text{-u.l.}}_{\text{DD}, j} \, (m_\chi)  \equiv \frac{\varmathbb{N}_\text{j}(m_\chi)}{N^{n_\sigma\text{-u.l.}}_j}  \;,
&& \varmathbb{A}^{n_\sigma\text{-u.l.}}_{\chi\chi\rightarrow \text{final}} \, (m_\chi)  \equiv \frac{\varmathbb{C}_\text{j}(\chi\chi\rightarrow \text{final};m_\chi) }{C^{n_\sigma\text{-u.l.}}_j(\chi\chi\rightarrow \text{final};m_\chi)}  \;.
\end{align}
With this notation, Eq.~(\ref{eq:ll-DAMA}), can be cast as
\begin{equation}
{\bf c}^T \; \varmathbb{A}^{n_\sigma\text{-l.l.}}_{\text{DAMA}[E_-,E_+]}(m_\chi)  \;{\bf c}> 1 \;, \\
\label{eq:out-ellipsoid}
\end{equation} 
while Eqs.~(\ref{eq:ul-DAMA}), (\ref{eq:ul-DD}), (\ref{eq:ul-SK}) can be collectively cast as
\begin{equation}
{\bf c}^T \; \varmathbb{A}^{n_\sigma\text{-u.l.}}_{j}(m_\chi) \;{\bf c}<1\;,
\label{eq:in-ellipsoid}
\end{equation}
with $j$ running over all the experimental upper limits. 

Geometrically, Eq.~(\ref{eq:out-ellipsoid}) corresponds to the exterior of an ellipsoid in a 28-dimensional parameter space, while Eq.~(\ref{eq:in-ellipsoid}) to the interior of an ellipsoid. Therefore, for a given dark matter mass $m_\chi$, the incompatibility  between the DAMA modulation signal in the bin $[E_-,E_+]$ and the null search results can be formulated by requiring an empty intersection between the region outside the ellipsoid represented by the $28\times 28$ matrix $\varmathbb{A}^{n_\sigma\text{-l.l.}}_{\text{DAMA}[E_-,E_+]}(m_\chi)$, which defines the region allowed by the DAMA experiment in the energy bin $[E_-,E_+]$,  and the regions enclosed by the ellipsoids represented by the matrices $ \varmathbb{A}^{n_\sigma\text{-u.l.}}_{j}(m_\chi)$, which define the region allowed  by the experimental upper limit $j$, with $j$ in a given set.  For the latter, we will consider the set consisting in the 3 upper limits on the modulation signal by DAMA, and the 9 upper limits from null search experiments listed in Tab.~\ref{tab:ul-ll}; we denote the set of these 12 experimental upper limits as ${\mathcal E}$. Furthermore, we will also consider the set $\mathcal{E}' $, consisting in $\mathcal{E}$ extended with the upper limit on the capture rate in the Sun from Super-Kamiokande.

\begin{figure}[t!]
\begin{center}
\hspace{-0.95cm}
\includegraphics[scale=0.92]{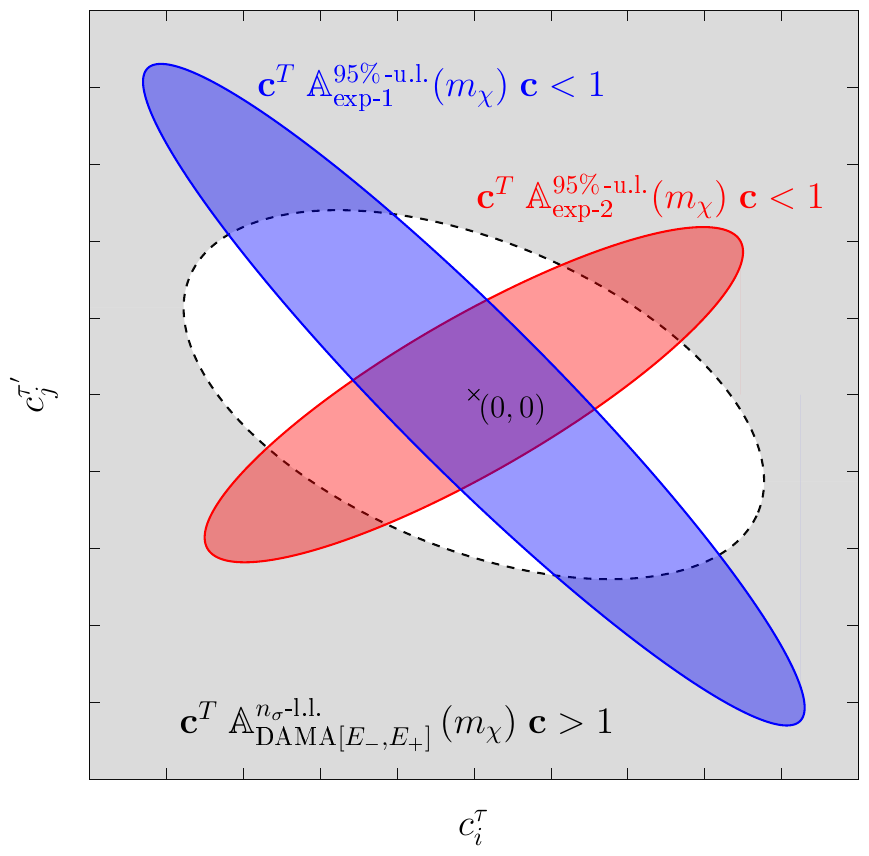}
\includegraphics[scale=0.92]{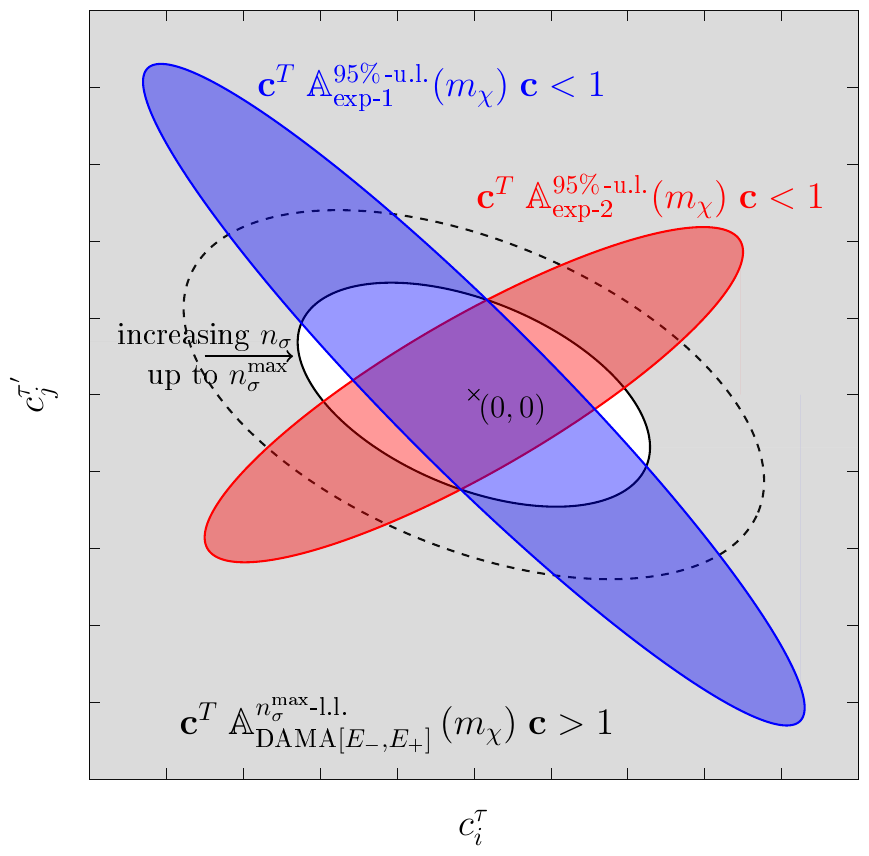}
\end{center}
\caption{Schematic description of our method to quantify the incompatibility of the DAMA modulation signal with the null search experiments. We show the case where the DAMA modulation signal is generated by the elastic scattering of a dark matter particle of mass $m_\chi$ with a nucleus, induced  for simplicity by only two operators $\hat{\mathcal{O}}_i$ and $\hat{\mathcal{O}}_j$ with coefficients $c_{i}^{\tau}$ and $c_{j}^{\tau^\prime}$. The regions of the parameter space allowed  at the 95\% C.L. by the experiments ``exp-1'' and ``exp-2'' are shown, respectively, as a blue and a red region (see. Eq.~(\ref{eq:in-ellipsoid})). On the other hand, the region required by the $n_\sigma$-lower limit on the modulation rate at DAMA is shown as a grey region (see Eq.~(\ref{eq:out-ellipsoid})). For a low value of $n_\sigma$, the overlap among the allowed regions of the three experiments is empty. Hence, following our prescription in the main text, we say that the DAMA modulation signal in the bin $[E_-,E_+]$ is inconsistent with the null results with a $n_\sigma$-significance (left panel). Increasing the value of  $n_\sigma$, the region allowed by DAMA accordingly increases, and eventually the intersection among the three allowed regions becomes non-empty (right plot). We then define $n^\text{max}_\sigma$ as the maximum value of $n_\sigma$ for which DAMA is inconsistent with the null results.}
\label{fig:procedure}
\end{figure}

More concretely, we consider the interior of the 95\% C.L. ellipsoids represented by the matrices $\varmathbb{A}^{95\%\text{-u.l.}}_{j}$ and we determine, for each of the three energy bins where the modulation signal has been reported, the maximum value of $n_\sigma$ such that the intersection with the ellipsoid  represented by $\varmathbb{A}^{n_\sigma\text{-l.l.}}_{\text{DAMA}[E_-,E_+]}(m_\chi)$ is empty;  this procedure is sketched in Fig.~\ref{fig:procedure}. We denote as  $n_\sigma^\text{max, (a)}$  the maximum value of  $n_\sigma$ in the energy bin $a=1,2,3$,  and, finally, we quantify the tension between the dark matter interpretation of the  DAMA modulation signal and the null search experiments by taking the maximum among these three values, $N_\sigma^\text{max}=\text{max}_{a \in \{1,2,3\}}\{n_\sigma^\text{max, (a)}\}$; the larger $N_\sigma^\text{max}$, the stronger the tension. We note that, for simplicity, we keep fixed the volume of the parameter space allowed by the null result experiments  by imposing a 95\% C.L. exclusion limit for all of them. Nevertheless, our procedure can be straightforwardly generalized to other choices of the exclusion significance, which may be different depending on the experiment. In particular, and in order to assess the robustness of our conclusions, we will also study the case when the region allowed by the null search experiments is enlarged by imposing a 99.9\% C.L. exclusion limit. 

To determine whether the intersection of the ellipsoids represented by the matrices $\varmathbb{A}^{95\%\text{-u.l.}}_{j}$, $j\in \mathcal{E}$, is fully contained within the  ellipsoid represented by $ \varmathbb{A}^{n_\sigma\text{-l.l.}}_{\text{DAMA}[E_-,E_+]}$, it is sufficient to find a set of real parameters $ \zeta_j \geq 0$ satisfying~\cite{Polik:2007:SS:1655208.1655210}
\begin{align}
&{\it i})\, \, \displaystyle{\sum_{j \,\in\, \mathcal{E}} \zeta_j < 1 } \, \,  \text{and} \nonumber \\
&{\it ii}) \, \, \displaystyle{\left( \sum_{j\, \in\, \mathcal{E}}  \zeta_j \,\varmathbb{A}^{95\%\text{-u.l.}}_{j} \right) - \varmathbb{A}^{n_\sigma \text{-l.l.}}_{\text{DAMA}[E_-,E_+]}} \,\, \text{is a positive definite matrix.}
\label{eq:zeta_condition}
\end{align}
If this is the case, there is no solution to Eqs.~(\ref{eq:out-ellipsoid}) and (\ref{eq:in-ellipsoid}), with $j\in\mathcal{E}$, and the DAMA modulation signal in the bin $[E_-,E_+]$ is incompatible with the null results\footnote{
Indeed, it is straightforward to check that
\begin{equation}
{\bf c}^T \; \varmathbb{A}^{n_\sigma\text{-l.l.}}_{\text{DAMA}[E_-,E_+]} {\bf c}<
{\bf c}^T \left( \sum_{j\, \in\, \mathcal{E}} \zeta_j \varmathbb{A}^{95\% \text{-u.l.}}_{j} \right){\bf c}<
\sum_{j\, \in\, \mathcal{E}} \zeta_j <1 \,,
\end{equation}
where in the first step we have used condition {\it ii}), in the second, eq.~(\ref{eq:in-ellipsoid}), and in the last, condition {\it i}).}. Using this procedure, we determine, for a given dark matter mass and energy bin $a$, the maximal significance $n_\sigma^{\text{max,(a)}}$ for which a set $\zeta_j$ satisfying the above conditions exists\footnote{For the numerical implementation of the algorithm, we used the {\tt feasp} solver implemented in MATLAB~\cite{MATLAB:2014a}.}, and finally, we construct the parameter $N_\sigma^\text{max}$, as defined above.

\section{Numerical results}
\label{sec:NumRes}

\begin{figure}
\vspace{0.2cm}
\begin{center}
\hspace{-0.75cm}
\includegraphics[scale=0.89]{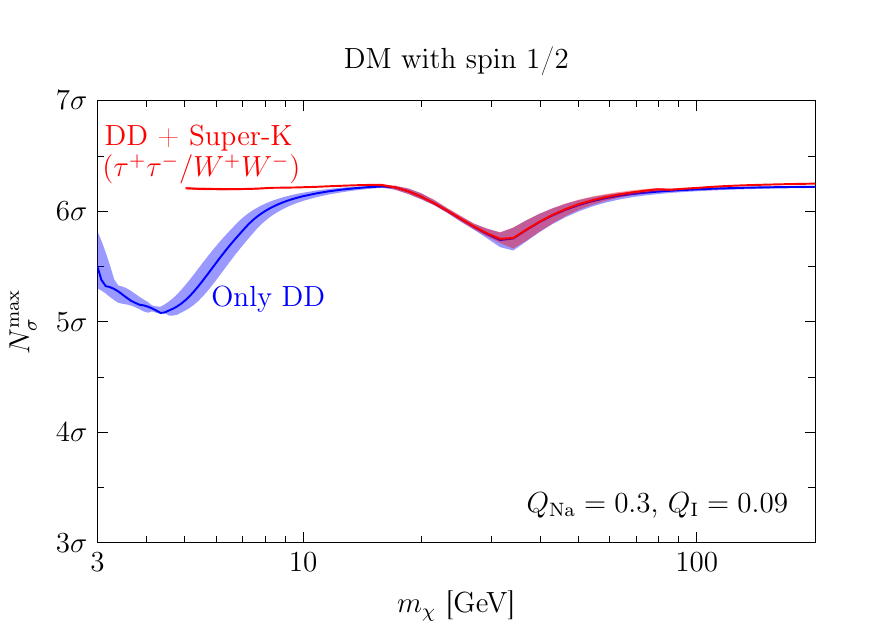}
\includegraphics[scale=0.89]{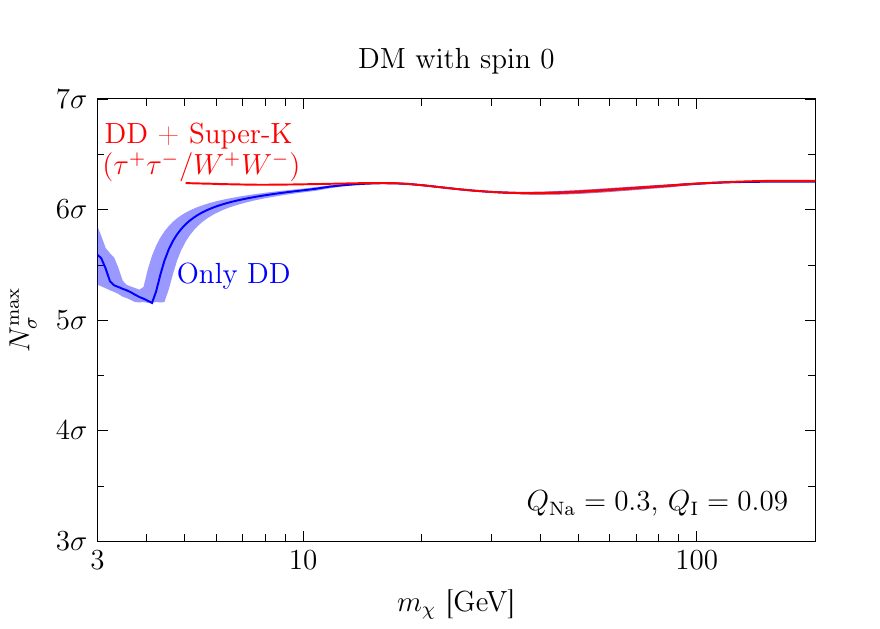} \\
\hspace{-0.75cm}
\includegraphics[scale=0.89]{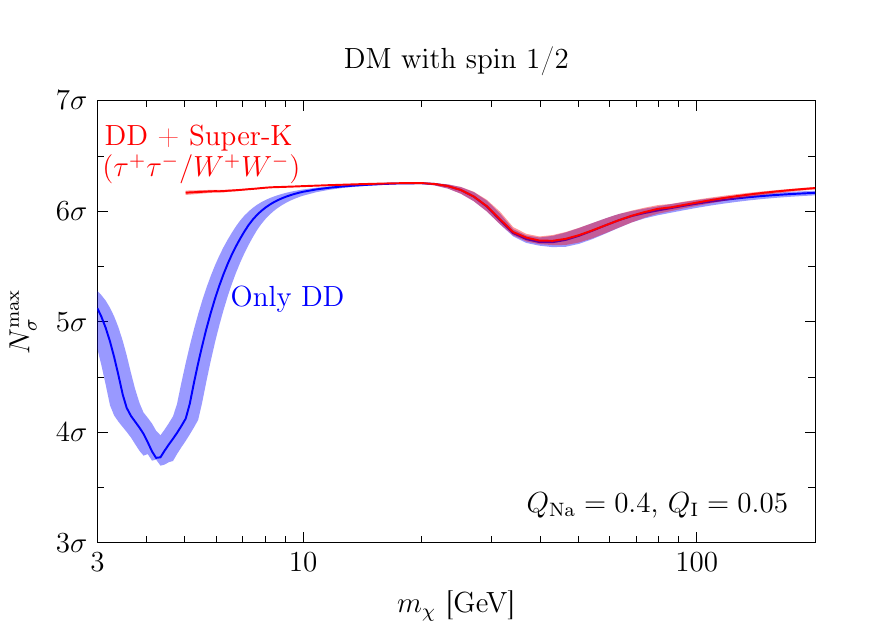}
\includegraphics[scale=0.89]{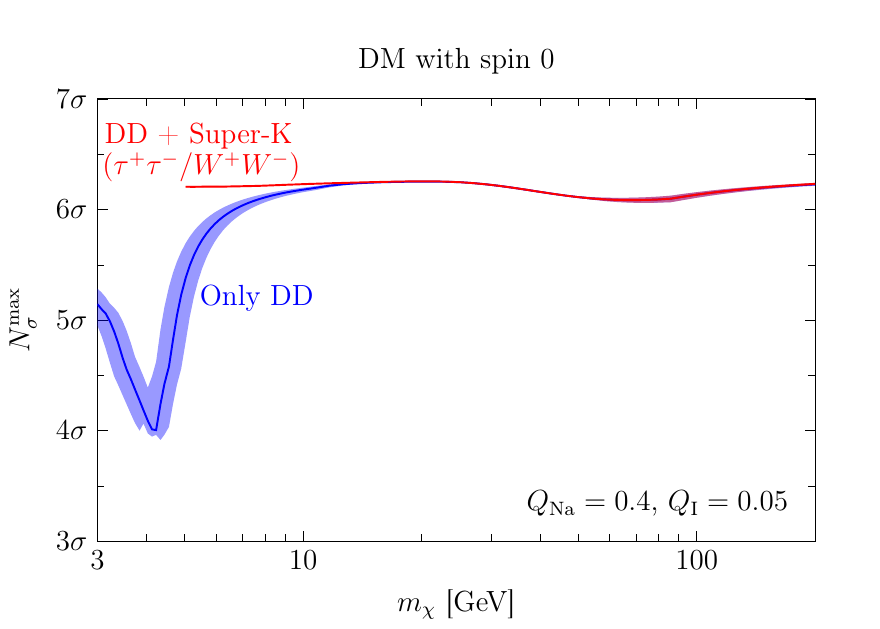}
\end{center}
\caption{Value of the parameter $N_\sigma^\text{max}$, which measures the tension between the DAMA modulation signal and the null search experiments, as a function of the dark matter mass $m_\chi$, for dark matter with spin 1/2 (left panels) or with spin 0 (right panels), assuming the set of quenching factors $Q_\text{Na}=0.3$ and $Q_\text{I}=0.09$ (upper plots) and $Q_\text{Na}=0.4$ and $Q_\text{I}=0.05$ (lower plots). The blue line was derived considering upper limits from direct detection experiments, while the red line also includes the upper limit on the capture rate in the Sun from Super-Kamiokande; the shaded bands bracket the uncertainties in the parameters of the Maxwell-Boltzmann dark matter velocity distribution.}
\label{fig:MaxSignal}
\end{figure}

We show in Fig.~\ref{fig:MaxSignal} the value of $N_\sigma^\text{max}$ as a function of the dark matter mass, for the case of a dark matter particle with spin 1/2 (left plots) or spin 0 (right plots); in the latter case, only the subset of operators not involving the dark matter spin, {\it  i.e.} $\left\{ \mathcal{O}_1, \mathcal{O}_3, \mathcal{O}_7, \mathcal{O}_\text{10} \right\}$, contribute to the scattering amplitude. Furthermore, we consider two choices of quenching factors $Q_{\rm Na}=0.3$ and $Q_{\rm I}=0.09$~\cite{Bernabei:1996vj} (upper plots), and $Q_{\rm Na}=0.4$ and $Q_{\rm I}=0.05$~\cite{Fushimi:1993nq} (lower plots). The blue curves were obtained by  taking into account only upper limits from direct detection experiments, \emph{i.e.}~using the set of upper limits $\mathcal{E}$ introduced in Section~\ref{sec:Confrontation}, while the red curves also include the null searches for a high energy neutrino flux from the Sun by Super-Kamiokande, namely the set $\mathcal{E}'$. We only show the results assuming dark matter annihilation into $ \tau^+\tau^-/W^+W^-$, since for annihilation into  $b\bar b$ the resulting values for $N_\sigma^\text{max}$ are practically identical to those obtained considering just the set $\mathcal{E}$. The solid curves were constructed using $v_0 = 230$ km/s and $v_\text{esc} = 533$ km/s in the Maxwell-Boltzmann velocity distribution, with the shaded bands bracketing the dependence of the result on changing $v_0$ within the range $220 - 240$ km/s and $v_\text{esc}$ within the range $492 - 587$ km/s. As apparent from the Figure, the value of $N_\sigma^\text{max}$ is only mildly sensitive to the choice of the velocity parameters of the Maxwell-Boltzmann distribution.

We find a strong tension between the  upper limits from all direct detection experiments and the lower limit on the modulation amplitude measured by DAMA in at least one of the energy bins. More precisely, we find $N_\sigma^\text{max}\gtrsim 5.1\sigma$, for quenching factors $Q_{\rm Na}=0.3$ and $Q_{\rm I}=0.09$, and $N_\sigma^\text{max}\gtrsim 3.7 \sigma$, for $Q_{\rm Na}=0.4$ and $Q_{\rm I}=0.05$. To assess the robustness of this conclusion, we have also calculated the value of $N_\sigma^\text{max}$ using the ellipsoids defined by the 99.9\% C.L. upper limits from the null search experiments. Namely we have replaced $\varmathbb{A}^{95\%\text{-u.l.}}_{j}$  by $\varmathbb{A}^{99.9\%\text{-u.l.}}_{j}$ in the procedure described in Section \ref{sec:Confrontation}  to calculate $N_\sigma^\text{max}$. As expected, we find a milder tension between DAMA and the null search experiments, although still significant.  Concretely, we find $N_\sigma^\text{max}\gtrsim 4.6\sigma~(3.2\sigma)$  for $Q_{\rm Na}=0.3$ and $Q_{\rm I}=0.09$ ($Q_{\rm Na}=0.4$ and $Q_{\rm I}=0.05$). Furthermore, when the set of upper limits from direct detection experiments $\mathcal{E}$ is extended with Super-Kamiokande, the tension with the DAMA modulation signal becomes, for $m_\chi\gtrsim 5$ GeV, more acute (for  $m_\chi\lesssim 5$ GeV evaporation effects in the Sun are significant and the Super-Kamiokande limits become irrelevant).

It is important to note that there is at present no consensus in the literature about the value of the quenching factors for Na and I in the DAMA crystals. This uncertainty is critical for establishing the incompatibility between DAMA and the null search experiments: since the measured modulation amplitude favors scatterings off Na, the tension among experiments gets milder as $Q_{\rm Na}$ increases. On the other hand, recent studies using low-energy pulsed neutrons claim $Q_{\rm Na}\lesssim 0.2$ in the range from 3 to 52 keV within a few per cent error~\cite{Xu:2015wha}, thus disfavoring values of $Q_{\rm Na}$ larger than 0.4. Should this upper limit be also applicable to the DAMA crystals, our conclusions would be strengthened.

\begin{figure}
\vspace{1.2cm}
\begin{center}
\hspace{-0.7cm}
\includegraphics[scale=0.89]{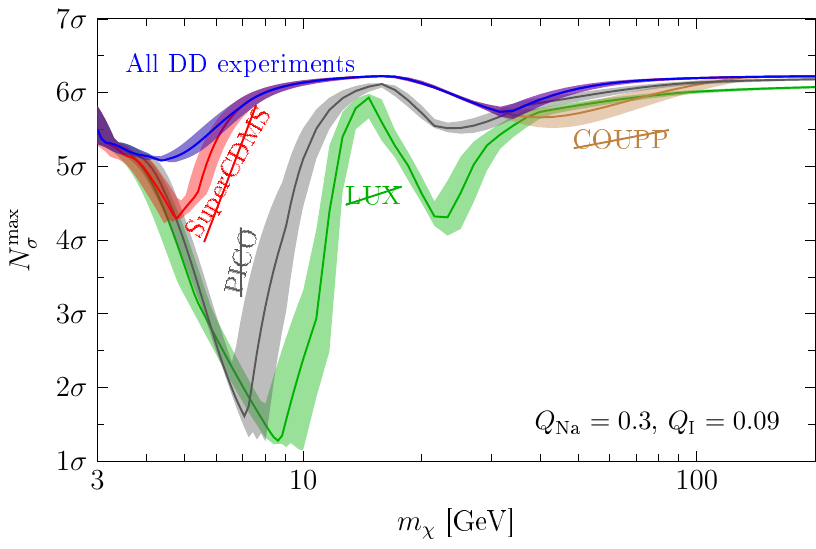}
\hspace{0.6cm}
\includegraphics[scale=0.89]{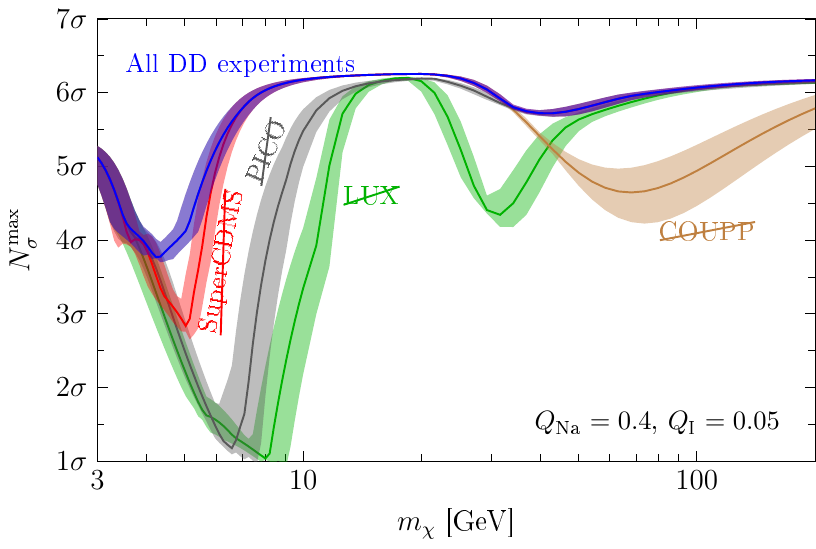}
\end{center}
\caption{Value of $N_\sigma^\text{max}$ as a function of the dark matter mass calculated excluding from the analysis one single direct detection experiment at a time, either Super-CDMS (red line), PICO (grey line), LUX (green line) or COUPP (brown line) for the quenching factors $Q_{\rm Na}=0.3$ and $Q_{\rm I}=0.09$ (left plot) and $Q_{\rm Na}=0.4$ and $Q_{\rm I}=0.05$ (right plot). The blue line shows for comparison the value of $N_\sigma^\text{max}$ calculated including all the upper limits from direct detection experiments.}
\label{fig:MaxSignal_QSmall_TakingOutExps}
\end{figure}

Finally, we investigate which of the null results in Table \ref{tab:ul-ll} is most relevant in testing the dark matter interpretation of the DAMA modulation signal. To this end, we have calculated $N_\sigma^\text{max}$ excluding one direct detection experiment from the set $\mathcal{E}$. The result is shown in Fig.~\ref{fig:MaxSignal_QSmall_TakingOutExps}, when excluding the Super-CDMS, PICO, LUX or COUPP results, compared to the value of $N_\sigma^\text{max}$ which follows from considering the full set $\mathcal{E}$. Notably, when excluding only the PICO results, or only the LUX results, we find that DAMA is well compatible with all other null searches. Geometrically, this result can be interpreted from the different orientations of the ellipsoids represented by $\varmathbb{A}^{95\%\text{-l.l.}}_\text{LUX}$ and $\varmathbb{A}^{95\%\text{-l.l.}}_\text{PICO}$, as illustrated in Fig.~\ref{fig:procedure}. As apparent from the figure, the volume enclosed by the intersection of both ellipsoids (which corresponds to the region of the parameter space allowed by both experiments) is much smaller than the volume enclosed by just one ellipsoid (which corresponds to the region of the parameter space allowed by a single experiment). Therefore, in the former case it is necessary a larger value of $n_\sigma$ in order to obtain a non-empty intersection with the exterior of the ellipsoid represented by $\varmathbb{A}^{n_\sigma\text{-l.l}}_{\text{DAMA}[E_-,E_+]}$ (which corresponds to the region allowed by the DAMA modulation signal in the bin $[E_-,E_+]$). This observation vindicates the necessity of employing various target nuclei in direct dark matter searches.

\section{Conclusions}
\label{sec:Conclusions}

We have investigated the compatibility of the modulation signal reported by the DAMA experiment with the null results from LUX, SuperCDMS, SIMPLE, PICO, COUPP, PICASSO and Super-Kamiokande, under the assumption that the dark matter scatters elastically off nuclei and that the velocity distribution in the Solar System follows a Maxwell-Boltzmann distribution. We have extended previous works by analyzing the scattering induced by the most general Galilean-invariant effective interaction of a spin 0 or 1/2 dark matter particle, taking into account possible interferences among operators. 

In order to exhaustively explore the 28-dimensional parameter space of the model, we have developed a novel method that allows to efficiently test whether there exists a non-empty intersection between the regions of the parameter space allowed by DAMA and the regions allowed by null searches. We have confronted the DAMA signal to the null results from direct detection experiments, considering for concreteness upper limits at the 95\% C.L. for all direct detection experiments. We have found that the dark matter interpretation of the DAMA modulation signal is excluded, in at least one energy bin where the signal has been reported, by more than $5.1\sigma$ in the whole range of dark matter masses, for quenching factors $Q_{\rm Na}=0.3$ and $Q_{\rm I}=0.09$, and by more than  $3.7\sigma$, for $Q_{\rm Na}=0.4$ and $Q_{\rm I}=0.05$. Including the 95\% C.L. upper limit on the dark matter capture rate in the Sun from Super-Kamiokande worsens the tension for dark matter masses above $\sim 5$ GeV. We have also checked the robustness of our conclusions by increasing the significance of the null results to the 99.9\% C.L. Finally, if the quenching factor for sodium recoils in the DAMA crystals turns out to be smaller than $Q_{\rm Na}<0.2$, as suggested by some recent studies, the tension between DAMA and the null result experiments would become even more acute. 

Some scenarios which are not covered by our analysis and where our conclusions may not hold are the following: {\it i)} dark matter particles with spin 1 or larger, leading to the presence of additional non-relativistic operators,~\emph{c.f.}~\cite{Dent:2015zpa}, {\it ii}) dark matter-nucleon interactions mediated by particles lighter than the typical momentum transferred in the scattering, {\it iii}) scenarios where the truncation in the effective theory expansion to operators at most quadratic in $\hat{\bf q}$ and linear in $\hat{{\bf v}}^\perp$ provides a poor approximation to the full Hamiltonian, {\it iv}) scenarios with inelastic dark matter-nucleus scattering~\cite{Barello:2014uda}, {\it v}) scenarios with two-body dark matter nucleon interactions~\cite{Hoferichter:2015ipa}, WIMP-electron and WIMP-atom scattering~\cite{Kopp:2009et}, or scenarios where the relativistic products of WIMP annihilations scatter off the target nuclei~\cite{Cherry:2015oca}, and {\it vi}) dark matter velocity distributions drastically different from the Maxwell-Boltzmann one. Further investigations in these directions, and most notably new dark matter searches employing also NaI crystals, as in the experiments DM-Ice~\cite{deSouza:2016fxg} or ANAIS~\cite{Amare:2016fmp}, will hopefully shed light on the origin of the DAMA modulation signal.

\section*{Acknowledgements}

The work of AI and SW has been partially supported by the DFG cluster of excellence ``Origin and Structure of the Universe'', the TUM Graduate School and the Studienstiftung des Deutschen Volkes. We thank Henrique Araujo and Felix Kahlhoefer for useful discussions regarding the LUX data.

\appendix

\section{Dark matter response functions}
\label{sec:appDM}
Using the same notation adopted in the body of the paper, we list below the dark matter response functions that appear in Eq.~(\ref{eq:sigma}). Notice that $v_T^{\perp 2}=v^2-q^2/(4 \mu_T^2)$, where $\mu_T$ is the dark matter-nucleus reduced mass, and $j_{\chi}$ is the dark matter particle spin:
\allowdisplaybreaks
\begin{eqnarray}
 R_{M}^{\tau \tau^\prime}\left(v_T^{\perp 2}, {q^2 \over m_N^2}\right) &=& c_1^\tau c_1^{\tau^\prime } + {j_\chi (j_\chi+1) \over 3} \left[ {q^2 \over m_N^2} v_T^{\perp 2} c_5^\tau c_5^{\tau^\prime }+v_T^{\perp 2}c_8^\tau c_8^{\tau^\prime }
+ {q^2 \over m_N^2} c_{11}^\tau c_{11}^{\tau^\prime } \right] \nonumber \\
 R_{\Phi^{\prime \prime}}^{\tau \tau^\prime}\left(v_T^{\perp 2}, {q^2 \over m_N^2}\right) &=& {q^2 \over 4 m_N^2} c_3^\tau c_3^{\tau^\prime } + {j_\chi (j_\chi+1) \over 12} \left( c_{12}^\tau-{q^2 \over m_N^2} c_{15}^\tau\right) \left( c_{12}^{\tau^\prime }-{q^2 \over m_N^2}c_{15}^{\tau^\prime} \right)  \nonumber \\
 R_{\Phi^{\prime \prime} M}^{\tau \tau^\prime}\left(v_T^{\perp 2}, {q^2 \over m_N^2}\right) &=&  c_3^\tau c_1^{\tau^\prime } + {j_\chi (j_\chi+1) \over 3} \left( c_{12}^\tau -{q^2 \over m_N^2} c_{15}^\tau \right) c_{11}^{\tau^\prime } \nonumber \\
  R_{\tilde{\Phi}^\prime}^{\tau \tau^\prime}\left(v_T^{\perp 2}, {q^2 \over m_N^2}\right) &=&{j_\chi (j_\chi+1) \over 12} \left[ c_{12}^\tau c_{12}^{\tau^\prime }+{q^2 \over m_N^2}  c_{13}^\tau c_{13}^{\tau^\prime}  \right] \nonumber \\
   R_{\Sigma^{\prime \prime}}^{\tau \tau^\prime}\left(v_T^{\perp 2}, {q^2 \over m_N^2}\right)  &=&{q^2 \over 4 m_N^2} c_{10}^\tau  c_{10}^{\tau^\prime } +
  {j_\chi (j_\chi+1) \over 12} \left[ c_4^\tau c_4^{\tau^\prime} + \right.  \nonumber \\
 && \left. {q^2 \over m_N^2} ( c_4^\tau c_6^{\tau^\prime }+c_6^\tau c_4^{\tau^\prime })+
 {q^4 \over m_N^4} c_{6}^\tau c_{6}^{\tau^\prime } +v_T^{\perp 2} c_{12}^\tau c_{12}^{\tau^\prime }+{q^2 \over m_N^2} v_T^{\perp 2} c_{13}^\tau c_{13}^{\tau^\prime } \right] \nonumber \\
    R_{\Sigma^\prime}^{\tau \tau^\prime}\left(v_T^{\perp 2}, {q^2 \over m_N^2}\right)  &=&{1 \over 8} \left[ {q^2 \over  m_N^2}  v_T^{\perp 2} c_{3}^\tau  c_{3}^{\tau^\prime } + v_T^{\perp 2}  c_{7}^\tau  c_{7}^{\tau^\prime }  \right]
       + {j_\chi (j_\chi+1) \over 12} \left[ c_4^\tau c_4^{\tau^\prime} +  \right.\nonumber \\
       &&\left. {q^2 \over m_N^2} c_9^\tau c_9^{\tau^\prime }+{v_T^{\perp 2} \over 2} \left(c_{12}^\tau-{q^2 \over m_N^2}c_{15}^\tau \right) \left( c_{12}^{\tau^\prime }-{q^2 \over m_N^2}c_{15}^{\tau \prime} \right) +{q^2 \over 2 m_N^2} v_T^{\perp 2}  c_{14}^\tau c_{14}^{\tau^\prime } \right] \nonumber \\
     R_{\Delta}^{\tau \tau^\prime}\left(v_T^{\perp 2}, {q^2 \over m_N^2}\right)&=&  {j_\chi (j_\chi+1) \over 3} \left[ {q^2 \over m_N^2} c_{5}^\tau c_{5}^{\tau^\prime }+ c_{8}^\tau c_{8}^{\tau^\prime } \right] \nonumber \\
 R_{\Delta \Sigma^\prime}^{\tau \tau^\prime}\left(v_T^{\perp 2}, {q^2 \over m_N^2}\right)&=& {j_\chi (j_\chi+1) \over 3} \left[c_{5}^\tau c_{4}^{\tau^\prime }-c_8^\tau c_9^{\tau^\prime} \right].
 \label{eq:R}
\end{eqnarray}

\section{Analysis of the direct detection experiments}
\label{sec:DD_analysis_details}

For the implementation of the results of the various direct detection experiments employed in this work, we follow the prescriptions suggested by the respective experimental collaborations as closely as possible. In particular, for each experiment reporting a null result, we validate our approach by comparing the published upper limits on the standard spin-independent and/or spin-dependent scattering cross section of dark matter with the upper limits derived using our framework. In each case, our upper limit either matches the results obtained by the corresponding experimental collaboration, or is slightly more conservative.

We employ the most recent data release of the \textbf{LUX} experiment~\cite{Akerib:2015rjg}, which is based on an exposure of $1.4 \cdot 10^4 \, \text{kg}\cdot\text{days}$. The detection efficiency is taken from Fig.~1 of~\cite{Akerib:2015rjg}; following the collaboration, we only take into account recoil energies above 1.1 keV. We define everything below the red solid curve in Fig.~2 of~\cite{Akerib:2015rjg} as the signal region, and assume that this corresponds to an additional factor 1/2 in the efficiency. By construction, this assumption is well-motivated for large dark matter masses, while it underestimates the number of signal events in the low mass region, leading to more conservative upper limits. Finally, we multiply the efficiency by an additional factor $(18/20)^2$, as we are only interested in the events passing the more stringent cut $r < $ 18 cm on the fiducial radius. Based on one observed event in the corresponding signal region, and not making any assumptions about the background, the 95\% (99.9\%) C.L. upper limit on the number of expected recoil events is $N_\text{max} = 4.74\,(9.23)$.

For \textbf{SuperCDMS}, the latest data release~\cite{Agnese:2014aze} is based on a run of 577 $\text{kg}\cdot\text{days}$ exposure using a Germanium target. We consider the recoil energy range between 1.6 and 10 keV, with the energy dependent efficiency given in Fig.~1 of~\cite{Agnese:2014aze}. The collaboration reported the observation of 11 events passing all the cuts, with a background level that is not fully understood. Hence, we conservatively set the background contribution to zero, leading to the 95\% (99.9\%) C.L. upper limit $N_\text{max} = 18.2\,(25.6)$.

The \textbf{SIMPLE} experiment is based on a C$_2 \,$Cl$\,$F$_5$ target; due to the absence of detailed nuclear form factor calculations for C and Cl, we only take into account the contribution from scattering off F, which leads to conservative upper limits. We use the combined Stage I and II data~\cite{Felizardo:2011uw}, with an exposure of 20.18 $\text{kg}\cdot\text{days}$. Following the collaboration, the efficiency is given by $\epsilon(E_R) = 1-\exp (-3.6 (E_R/E_T-1))$, with an energy threshold of $E_T=$ 8 keV. In total, 11 events were observed, with a conservative background expectation of 14.5 events~\cite{Felizardo:2011uw}. Employing the Feldman-Cousins method, we obtain $N_\text{max} = 5.26 \, (11.9)$ for the 95\% (99.9\%) C.L. upper limit.

\textbf{COUPP} is based on a C$\,$F$_3\,$I target, for which we take into account the scattering off F and I only. The latest result~\cite{Behnke:2012ys} consists of three non-overlapping data sets with nucleation thresholds $E_T$ of 7.8 keV, 11.0 keV, and 15.5 keV, respectively, which we treat as three independent experiments. In each case, the bubble efficiency is assumed to be $\epsilon(E_R) = 1- \exp \left( 0.15 \left( 1-E_R/E_T\right) \right)$ for scattering off F, and $\epsilon(E_R)= \theta \left( E_R - E_T\right)$ for I. The effective exposures are given by by 55.8 kg $\cdot$ days, 70.0 kg $\cdot$ days, and 311.7 kg $\cdot$ days, respectively. Following the collaboration, we use the data including the time isolation cut, leading to 2, 3, and 8 observed events in the three different data sets. Again setting the background conservatively to zero, we obtain $N_\text{max}=6.30 \, (11.2)$, $7.75 \, (13.0)$, $14.4 \, (21.1)$, respectively, at the 95\% (99.9\%) C.L.

The \textbf{PICASSO} experiment uses a C$_4\,$F$_{10}$ target, and again we only take into account scattering off F. Among the various data sets presented in~\cite{Archambault:2012pm}, we only consider results obtained with energy thresholds of 1.73 keV and 2.9 keV, which turn out to be the most constraining ones for our purposes. The bubble efficiency is taken to be $1-\exp\left( 2.5 \left( 1-E_R/E_T\right)\right)$. Following the collaboration, we assume that the error in each energy bin is Gaussian distributed, and we calculate the 95\% (99.9\%) C.L. Feldman Cousin upper limits to be $8.72\,(18.0)$ events per kg of fluor per day for bin number 1, and $3.21\,(5.59)$ for bin number 2.

Finally, we employ the latest data from the \textbf{PICO} experiment, based on a C$_3\,$F$_{8}$ target with a total exposure after cuts of 129 kg$\,\cdot\,$days~\cite{Amole:2016pye}. We again neglect scattering off C, and use the black dashed line from Fig.~3 of~\cite{Amole:2015lsj} as the efficiency for scattering off F. One single-nuclear recoil event has been observed, corresponding to $N_\text{max} = 4.74\,(9.23)$ at 95\% (99.9\%) C.L., again assuming zero background. 

\bibliographystyle{JHEP-mod.bst}
\bibliography{ref}

\end{document}